\documentclass[CEJP,PDF]{cej} 

\title{Azimuthal correlations in Pb--Pb and pp collisions measured with the ALICE detector}

\articletype{Editorial}

\author{You Zhou\inst{1}$^,$\inst{2}\email{yzhou@nikhef.nl; you.zhou@cern.ch} 
(for the ALICE Collaboration)}

\institute{
     \inst{1} Nikhef, Science Park 105, 1098 XG Amsterdam, The Netherlands
     \inst{2} Utrecht University, P.O.Box 80000, 3508 TA Utrecht, The Netherlands}

\abstract{
We present results from the measurements of azimuthal correlations 
of charged particles in
$\sqrt{s_{_{NN}}}$ = 2.76 TeV Pb--Pb collisions and $\sqrt{s_{_{NN}}}$ = 7 TeV pp collisions. 
 In addition, the comparison of the experimental measurements in pp collisions
 with those from Pythia and Phojet simulations are presented.
}


\begin{document}
\maketitle

\section{Introduction}
The study of azimuthal correlations is one of the most important tools to 
probe the properties of the medium generated in heavy--ion collisions.
Experimentally, these azimuthal correlations are not
determined solely by anisotropic flow~\cite{JYO-PRD} but also have other contributions,
usually refered to as non--flow which are not correlated to the participant 
plane~\cite{ART-arXiv}.  
Anisotropic flow, especially the second order harmonic $v_{2}$ (elliptic flow), 
has been systematically studied from SPS to LHC energies~\cite{SPS-v2, RHIC-v2, LHC-v2}.
Recently it has been argued that fluctuations in the initial matter distribution
give rise to odd harmonics like $v_{3}$ (triangular flow)~\cite{BA-PRC}. 
In this contribution, we report the anisotropic flow for 
charged particles measured in $\sqrt{s_{_{NN}}} =$ 2.76 TeV Pb--Pb collisions. 
We also discuss azimuthal correlation measurements
in pp collisions compared to simulations from Pythia and Phojet.

\section{Anisotropic flow in Pb--Pb collisions}
\begin{figure}[thb]
\begin{center}
\includegraphics[width=11cm, height=6cm]{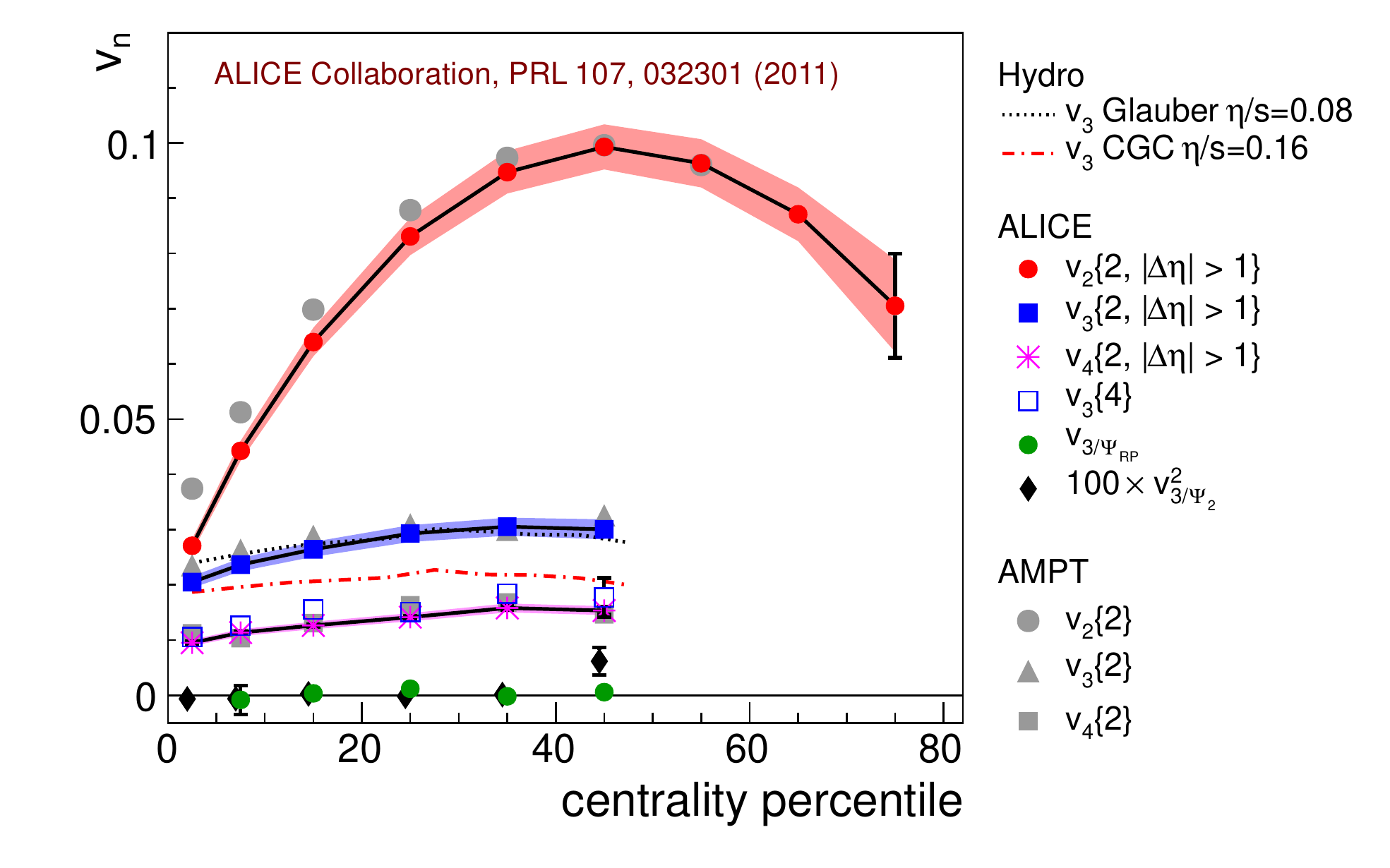}
\caption{$v_{2}$, $v_{3}$ and $v_{4}$ $p_t$-integrated flow as a function of centrality. 
Full and open blue squares show the $v_{3}\{2\}$ and $v_{3}\{4\}$, respectively. 
The full circle and full diamond are symbols for
$v_{3/\Psi_{\mathrm{RP}}}$ and $v^{2}_{3/\Psi_{2}} $. In addition,
 the hydrodynamic calculations~\cite{BHA-PRC}
for $v_{3}$ and AMPT simulations~\cite{JX-PRC} for $v_{2}$, $v_{3}$ and $v_{4}$ are shown by dash lines and full gray markers. 
ALICE data points taken from~\cite{ALICE-V3}.}
\label{Fig1}
\end{center}
\end{figure}

\begin{figure}[thb]
\includegraphics[width=5.cm, height=5.0cm]{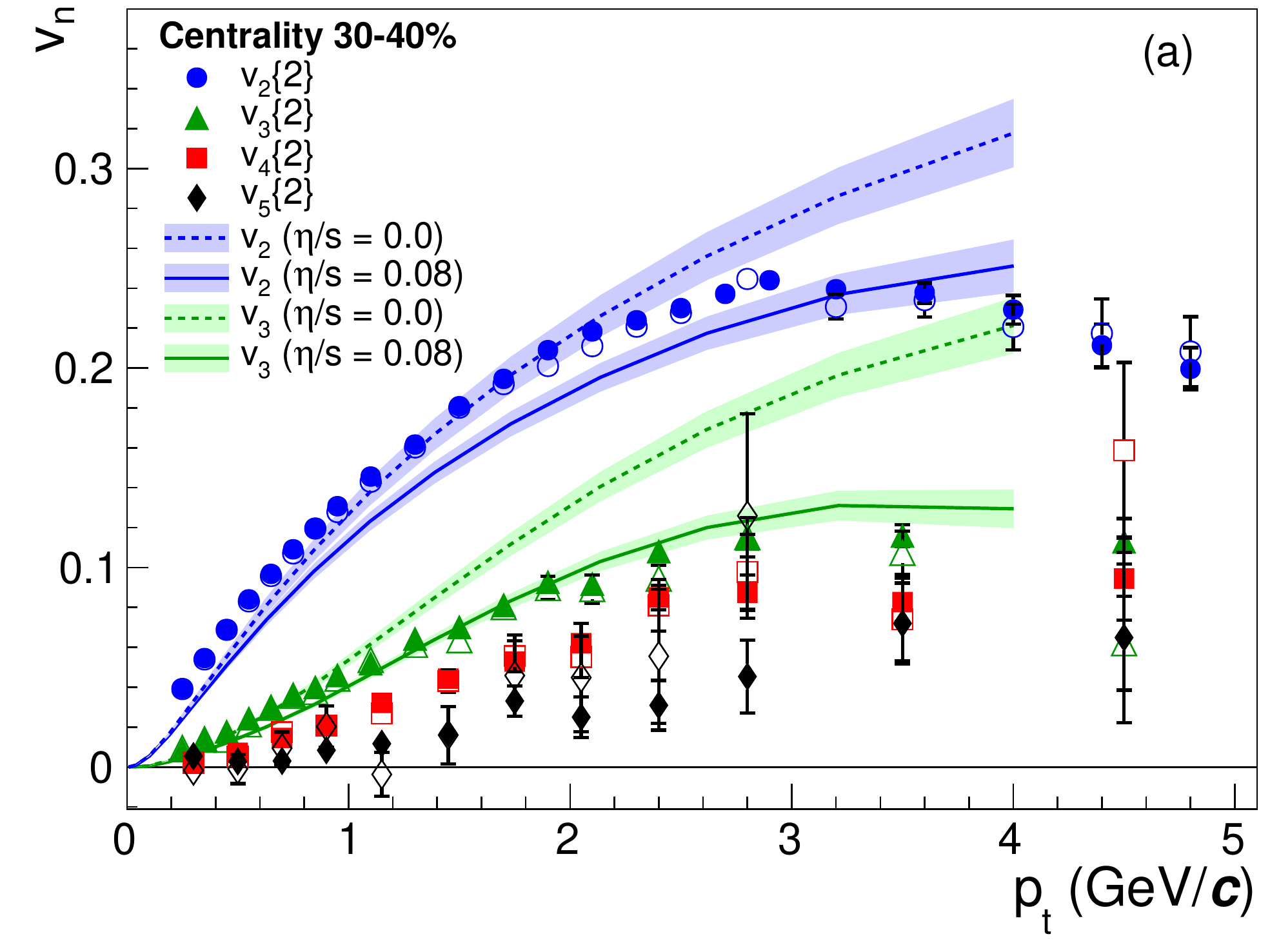}
\includegraphics[width=10.5cm, height=5.5cm]{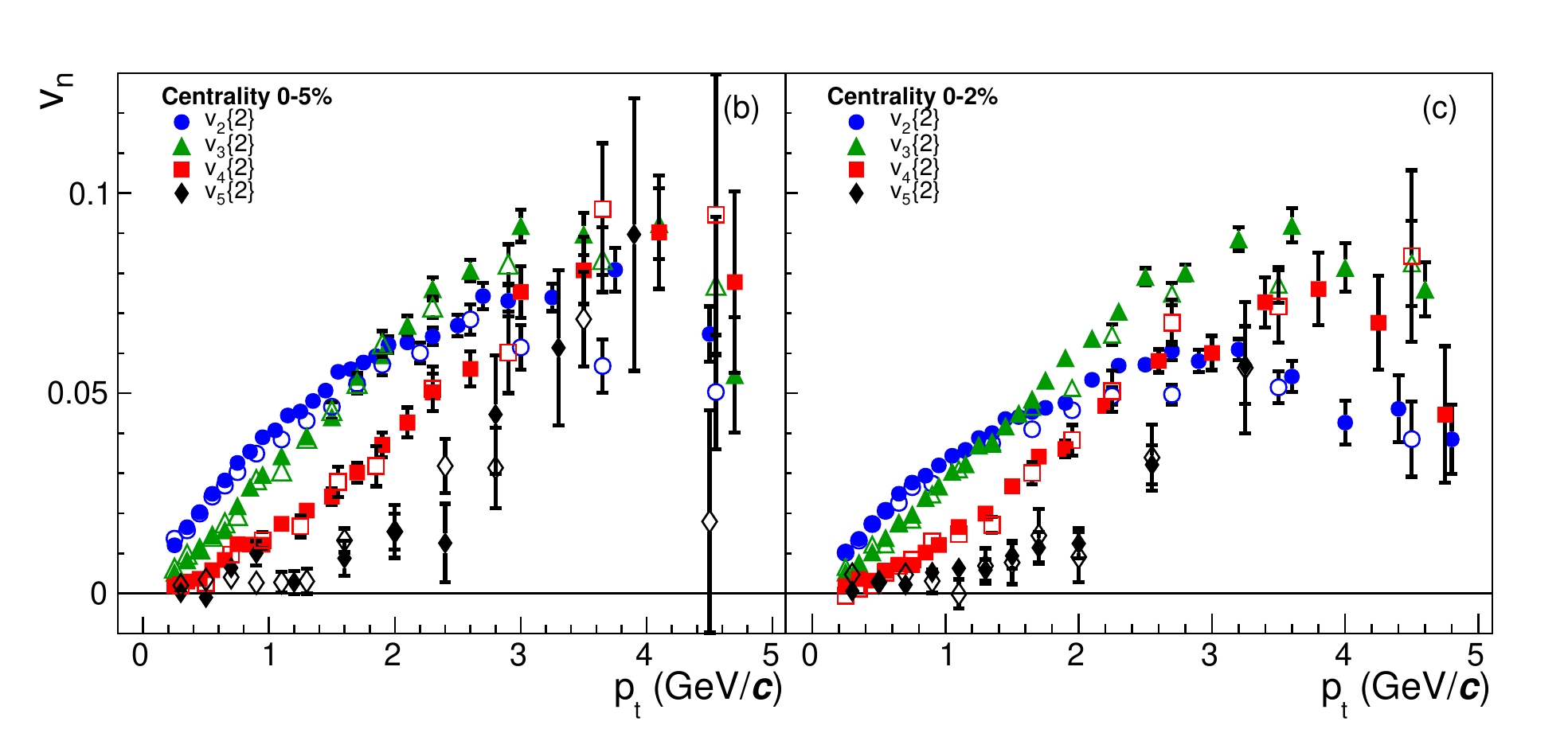}
\caption{$v_{2}$, $v_{3}$, $v_{4}$, $v_{5}$ as a function of transverse
momentum and for three event centralities. The full and open
symbols are for $|\Delta \eta|>$0.2 and $|\Delta \eta|>$1.0, 
respectively. (a) 30-40$\%$ centrality percentile compared to hydrodynamic
 model calculations~\cite{BS-arXiv}, (b) 0-5$\%$ centrality percentile, 
 (c) 0-2$\%$ centrality percentile. Figures taken from~\cite{ALICE-V3}.}
\label{Fig1}
\end{figure}

In this contribution, we report on the study the azimuthal correlations 
via 2-- and 4--particle
cumulants~\cite{AB-QC}.
In Fig. 1 we observe that the $v_{3}$ measurements from the 2-- and 4--particle
cumulants differ from zero;
the $v_{3}\{4\}$ is a factor of 2 smaller than $v_{3}\{2\}$ which can be understood 
  if $v_{3}$ originates predominantly from event--by--event fluctuations 
  of the initial spatial geometry~\cite{RSB-PRC}. 
 At the same time, we investigate the correlation between
 $\Psi_{3}$ and the reaction plane $\Psi_{\mathrm{RP}}$ as well as the 
 correlations between $\Psi_{3}$ and $\Psi_{2}$, evaluated by $v_{3/\Psi_{\mathrm{RP}}} 
  = \langle \cos ( 3 \phi - 3 \Psi_{\mathrm{RP}}) \rangle$ and
   $v_{3/\Psi_{2}}^{2} = \langle \cos (3 \phi_{1} + 3 \phi_{2} - 2 \phi_{3} - 
  2 \phi_{4} - 2 \phi_{5} )  \rangle / v_{2}^{3}$, respectively. 
  We observe that 
  $v_{3/\Psi_{\mathrm{RP}}}$ and $v_{3/\Psi_{2}}^{2}$
 are consistent with zero within uncertainties. Based on these results, 
 we conclude that $v_{3}$ develops as a correlation of all particles with respect 
 to the third order participant plane $\Psi_{3}$, while there is no (or very weak) 
 correlation between $\Psi_{\mathrm{RP}}$ (or $\Psi_{2}$) and $\Psi_{3}$.
The centrality dependence of $v_{3}$ is compared to hydrodynamic calculations.
The data are described well by calculations based on Glauber initial conditions 
and $\eta/s = 0.08$,
while underestimated by the MC--KLN initial conditions 
and $\eta/s = 0.16$~\cite{BHA-PRC}. 
 The comparison suggests that $\eta/s$ of the produced matter is small. 
 Finally, the data are described well by the AMPT model calculations, with
 only a slight overestimation of $v_{2}\{2\}$ in the most central collisions~\cite{JX-PRC}.  

To further constrain the properties of the system, we compare the $p_{t}$--differential
flow of $v_{2}$ and $v_{3}$ to hydrodynamic calculations in Fig. 2(a).
We find that the hydrodynamic
calculations with Glauber initial conditions can describe the elliptic and triangular
differential flow measurements, although not for higher $p_{t}$. 
However, the $v_{2}(p_{t})$ measurements seem to suggest $\eta/s$=0
while for $v_{3}(p_{t})$ the hydrodynamic calculations with $\eta/s$=0.08 provide
a better description. Currently there is no hydrodynamic calculation which simultaneously 
describes the $p_{t}$--differential $v_{2}$ and $v_{3}$ measurements 
at LHC energies with the same value for $\eta/s$.  
In central collisions 0-5$\%$ we observe that the higher harmonics $v_{3}$
and $v_{4}$ exceed $v_{2}$ and become the dominant harmonics 
at intermediate $p_{t}$.
This occurs already at lower $p_{t}$ for more central 
collisions 0-2$\%$. 
In AMPT simulations, it is observed that the initial geometrical fluctuations leads to 
anisotropic collective expansions even at an impact parameter of b=0~\cite{GL-PRL}.

\begin{figure}[thb]
\begin{center}
\includegraphics[width=14cm, height=6cm]{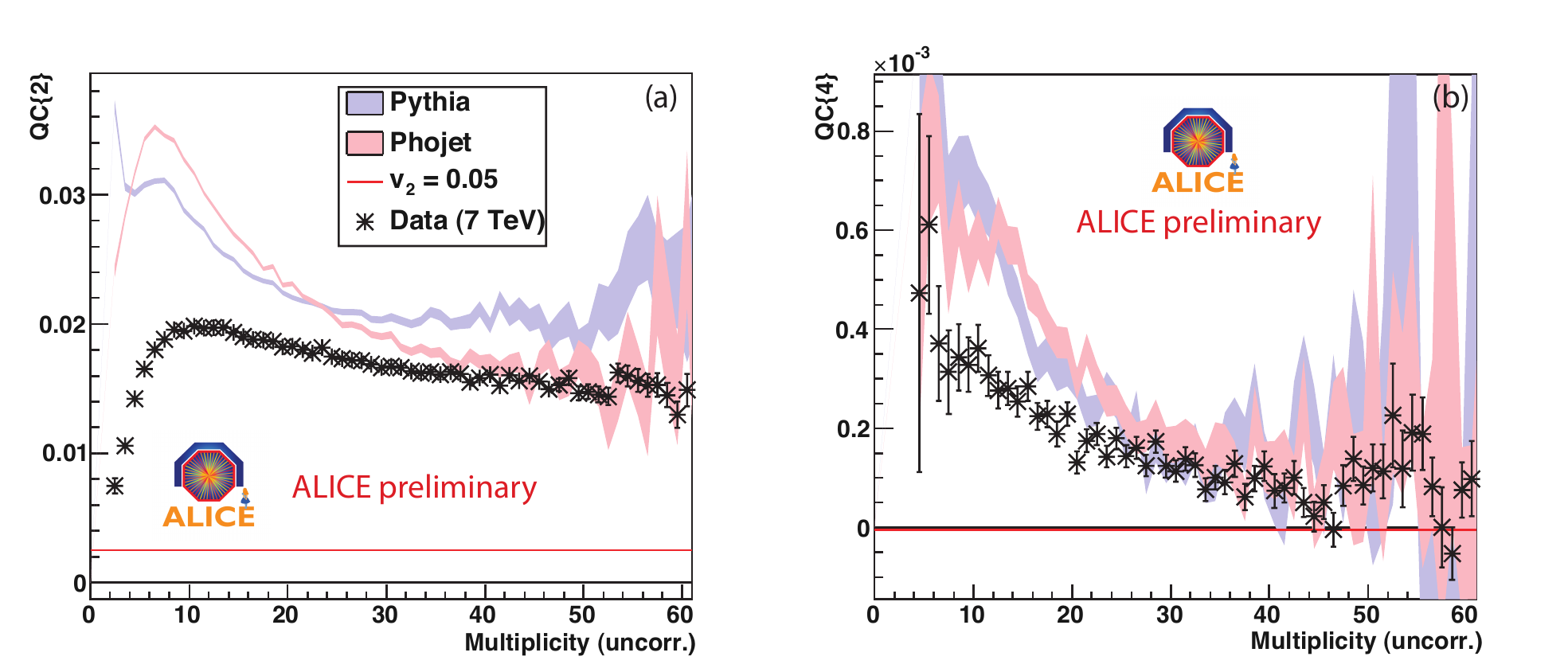}
\caption{Cumulants for charged particles in 7 TeV pp collisions. (a) 2-particle cumulant ; (b) 4-particle cumulant. The shadow areas represent the results for Pythia (purple) and Phojet (pink).}
\label{Fig1}
\end{center}
\end{figure}

\section{Anisotropic flow or non--flow in pp collisions?}
At LHC energies relatively high multiplicity events are observed in 
pp collisions~\cite{LHC-pp}. 
Some theoretical work predict elliptic flow 
magnitudes up to 0.2 in pp collisions at LHC energies~\cite{hydro-pp}.
It is interesting to investigate whether collective effects appear in such events 
and if we can test those predictions. The 2-- and 4--particle cumulant
when dominated by anisotropic flow, correspond to:
 QC$\{2\}=v^{2}$,~QC$\{4\}=-v^{4}$.
Therefore if the measured azimuthal correlations are dominated by anisotropic flow, 
they should show the typical flow signature (+,--) which has been 
observed in Pb--Pb collisions~\cite{AB-QM}.
Figure 3 presents the 2-- and 4--particle cumulant as 
a function of the measured
uncorrected multiplicity, defined as 
the number of charged particle tracks which pass our track selection.
We observe that the measured QC$\{4\}$ is positive 
in the currently measured multiplicity range, which suggests that its 
dominant contribution is not coming from anisotropic flow. 
Also we find that both QC$\{2\}$ and QC$\{4\}$ decrease with increasing multiplicity,
which is a typical behaviour for non--flow. In addition, we notice that both Pythia 
and Phojet can qualitatively describe the trend and sign of the QC$\{2\}$ and 
QC$\{4\}$. However, both of them do overestimate the strength of the 
azimuthal correlation measurements.

\section{Conclusion}
The azimuthal correlations of charged particles measured in $\sqrt{s_{_{NN}}}$
 = 2.76 TeV Pb--Pb collisions are presented. Our results constrain 
 the corresponding models. The analyses with 2-- and 4--particle cumulant
 in $\sqrt{s_{_{NN}}}$ = 7 TeV pp collisions show that such azimuthal correlations are not dominated by anisotropic flow in the multiplicity range presented.

\end{document}